\documentclass[11pt, a4paper, nofootinbib]{article}
\usepackage{graphicx}
\usepackage[bookmarks, colorlinks, breaklinks, pdfauthor={Huw Price}]{hyperref}
\usepackage{tikz}
\def\R2Lurl#1#2{\mbox{\href{#1}{\tt #2}}}

\usepackage{amsmath, amsthm, amssymb}

\usepackage{sectsty}
\subsubsectionfont{\hspace{-0.60cm}
\normalfont\emph}
\subsectionfont{\hspace{-0.60cm}\nohang\normalfont\normalsize\textit}
\sectionfont{\nohang\centering\normalfont\textsc}

\newcommand{\comment}[1]{}

\hypersetup{linkcolor=blue,citecolor=blue,filecolor=black,urlcolor=blue}

\def\str#1{{\setbox1=\hbox{#1}\leavevmode 
\raise.47ex\rlap{\hspace{4.5pt}\leaders\hrule\hskip\wd 1} 
\box1}}

               \usepackage{latexsym}

\DeclareRobustCommand\textprime{\leavevmode \raise.8ex\hbox{\text@char\scriptfont\prime}}

\begin{document}               
    \title{A live alternative to quantum spooks}
\author{Huw Price\thanks{Trinity College, Cambridge CB2 1TQ, UK; email \href{mailto:hp331@cam.ac.uk}{hp331@cam.ac.uk}.} {\ and} Ken Wharton\thanks{Department of Physics and Astronomy, San Jos\'{e} State University, San Jos\'{e}, CA 95192-0106, USA; email \href{mailto:kenneth.wharton@sjsu.edu}{kenneth.wharton@sjsu.edu}.}}
\date{}  

\maketitle
\begin{abstract}
\noindent Quantum weirdness has been in the news recently, thanks to an ingenious new experiment by a team led by Roland Hanson, at the Delft University of Technology. Much of the coverage presents the experiment as good (even conclusive) news for spooky action-at-a-distance, and bad news for local realism. We point out that this interpretation ignores an alternative, namely that the quantum world is retrocausal. We conjecture that this loophole is missed because it is confused for superdeterminism on one side, or action-at-a-distance itself on the other. We explain why it is different from these options, and why it has  clear advantages, in both cases. 
\end{abstract}

Quantum weirdness has been getting a lot of attention recently, thanks to a clever new experiment by a team led by Roland Hanson, at the Delft University of Technology.\footnote{Hanson, R.~\emph{et al, }`Experimental loophole-free violation of a Bell inequality using entangled electron spins separated by 1.3 km', \emph{Nature,} online 21 October 2015 [\href{http://www.nature.com/nature/journal/vaop/ncurrent/full/nature15759.html}{doi:10.1038/nature15759}][\href{http://arxiv.org/abs/1508.05949v1}{arXiv:1508.05949}].} Much of the coverage has presented the experiment as good news for spooky action-at-a-distance, and bad news for Einstein: ``The most rigorous test of quantum theory ever carried out has confirmed that the `spooky action-at-a-distance' that [Einstein] famously hated \ldots\ is an inherent part of the quantum world,'' as a report in \emph{Nature} put it.\footnote{Merali, Z., `Quantum `spookiness' passes toughest test yet', \emph{Nature,} 27 August 2015 [\href{http://www.nature.com/news/quantum-spookiness-passes-toughest-test-yet-1.18255}{http://www.nature.com/news/quantum-spookiness-passes-toughest-test-yet-1.18255}].}

Hanson's experiment does, as claimed, make a convincing case for closing some of the best-known loopholes in the case for action-at-a-distance. However, it is much further from settling the case against Einstein than these responses suggest. There's a large and promising loophole that simply doesn't show up on most commentators' radar -- or most physicists' radar, for that matter.\footnote{A commentary piece published in \emph{Nature} in conjunction with the Hanson \emph{et al} paper claims that the new findings ``rigorously reject'' the hypothesis of \emph{local realism} -- i.e., that the ``world is made up of real stuff, existing in space and changing only through local interactions.''  (Wiseman, H., `Quantum physics: Death by experiment for local realism', \emph{Nature,} 21 October 2015,  [\href{http://www.nature.com/nature/journal/vaop/ncurrent/full/nature15631.html}{doi:10.1038/nature15631}]) The same claim seems to be endorsed by Hanson \emph{et al} themselves, when they write: ``Our observation of a statistically significant loophole-free Bell inequality violation thus indicates rejection of all local-realist theories that accept that the number generators produce a free random bit in a timely manner and that the outputs are final once recorded in the electronics.'' However, the option we describe here is entirely compatible with local realism.} When it gets noticed at all, it tends to get confused for something else. So the fact that there's still a viable alternative to action-at-a-distance -- arguably, a much more attractive alternative, and certainly one that is untouched by the results from Hanson and his team -- remains a well-hidden secret.

This invisible loophole falls into a well-known category. The argument for action-at-a-distance assumes that quantum particles don't know what measurements they are going to encounter in the future. A little more technically, it assumes that the state of a particle before a measurement is independent of the particular setting chosen for that measurement (the choice whether to measure position or momentum, say). This sounds innocuous enough. How could the particle know about that, before it reaches the measurement device? But innocuous or not, it is crucial. Without this independence assumption, the argument for action-at-a-distance just doesn't go through.

The invisible loophole rejects this independence assumption, but it is confused for and obscured by another proposal for doing the same thing. This better-known cousin is a well-recognised but deservedly unpopular little loophole called \emph{superdeterminism.} To see how superdeterminism proposes to reject the independence assumption -- and how there's a much more attractive way of doing the same thing -- let's take a detour via medieval theology. (Superdeterminism has ancient ancestors.)

\subsection*{\textbf{\emph{Medieval interlude}}}
History (curated by us for this occasion) tells of the fast-talking fourteenth century theologian, Brother Bob, who was spotted one day throwing a stone through the stained-glass window of Abbess Alice. ``Oh, Brother,'' chideth the Abbess (something of a metaphysician herself), ``You have caused a stone to fly through the air, and that in turn has caused our precious window to break. (But I forgive you.)''

``Thank you, kind Abbess,'' replieth Bob, ``But there is surely nothing to forgive? Undoubtedly, the several movements of my arm, the stone, and your unfortunate window, were all caused by He who causes all things. It would be heresy -- indeed, Heresy with a capital H -- for me to take the blame.''

Brother Bob is what philosophers call an \emph{occasionalist.} He denies that there is any causation in the world -- at least, in the ordinary material world of arms, rocks, and windows. Instead, the appearance of worldly causation is all arranged by God. Whatever its theological merits, occasionalism has not survived the rise of modern science. Alice's description of the situation is clearly the right one, by scientific lights (as well as by ordinary lights). 

Yet occasionalism, or something very much like it, is the idea that superdeterminism proposes to resurrect, to violate the independence assumption. Brother Bob suggested that something in the background -- i.e., in this case, God -- explains the correlation between his arm movements, the stone's flight, and the broken window. Analogously, superdeterminists propose that something in the background -- some new kind of hidden variable, perhaps, or just some specific arrangement of the initial conditions of the Universe -- manages to coordinate the properties of quantum particles with choices of settings for measurements that those same particles are yet to encounter. This coordination makes it the case that particles with one property encounter position measurements, say, while particles with another property encounter momentum measurements. As a result, the independence assumption fails, and the argument for action-at-a-distance collapses. 

Opponents object that this coordination seems to conflict with our view that the measurement setting can be freely chosen, whether by a human experimenter or by some random device (such as the Swiss national lottery machine, as John Bell once suggested). This lack of free will was a virtue for Brother Bob -- it got him off the hook for breaking the window! But for modern physicists it seems a major objection to superdeterminism. Indeed, as opponents point out, the experimenter's freedom of choice seems presupposed by science. 

Still, superdeterminism does get a mention in discussion of the Hanson experiment. As \emph{New Scientist} notes, for example:
\begin{quotation}
\noindent There is one remaining loophole for local realists to cling to, but no experiment can ever rule it out. What if there is some kind of link between the random microwave generators and the detectors? Then Alice and Bob might think they're free to choose the settings on their equipment, but hidden variables could interfere with their choice and thwart the Bell test.\footnote{`Quantum weirdness proved real in first loophole-free experiment', \emph{New Scientist,} Daily News, 28 August 2015. [\href{https://www.newscientist.com/article/dn28112-quantum-weirdness-proved-real-in-first-loophole-free-experiment/}{https://www.newscientist.com/article/dn28112-quantum-weirdness-proved-real-in-first-loophole-free-experiment/}]}
\end{quotation}
Remarkably, however, all these discussions seem entirely blind to the fact that there's an option that sits right alongside superdeterminism, differing from it in exactly the way that the ordinary explanation of the broken window (that Bob caused the rock to fly, and so caused the window to break) differs from Bob's occasionalist proposal. This alternative version still blocks the argument for action-at-a-distance, by rejecting the independence assumption -- but, as in the fourteenth century, it saves free will. 

\subsection*{\textbf{\emph{Back to the future}}}

To explain this invisible alternative, let's fast-forward 700 years, and give the floor to modern Alice and her brother Bob, who are quantum physicists, and distant descendants of their medieval namesakes (whose friendship blossomed, after that ``rocky'' start -- they fell in love, left the church, and raised a family). Once again, Bob is the superdeterminist. ``I know it's an implausible explanation of these entanglement experiments,'' he says to Alice, ``But anything is better than spooky action-at-a-distance. Our famous ancestors rejected spooks seven centuries ago, as did the Great One (I mean Einstein) more recently. We should do the same.''

``I agree with you about the spooks, brother,'' says Alice, ``But isn't there a better alternative? Of course we are free to choose the measurement settings, but why not say that the prior properties of the particles are \emph{caused} by the measurement settings, and hence indirectly by our choices? It's just like ordinary causation, except that in the quantum world causation turns out to work backwards as well as forwards. \emph{Retrocausality} -- perhaps that's the real lesson of the quantum correlations!''

Alice's suggestion is the overlooked spook-busting loophole we promised you at the beginning. Mathematically speaking, it avoids the spooks in exactly the way that superdeterminism does, by giving up the assumption that the hidden properties of particles are independent of the measurements they are going to encounter in the future. But it saves free will, in just the way that the ordinary causal story about stones and windows saved free will, back in the fourteenth century. Alice and Bob freely choose measurement settings, just as their ancestor freely chose to throw that stone. The measurement settings cause the incoming particle to have certain properties, just as the stone caused the window to break. The only difference is that in this case, down at the microscopic level, some of the causation works backwards.

At this point, Bob may feel that the proposed cure is worse than the disease. Isn't it just \emph{obvious} that causation doesn't work backwards? Well, maybe, but why, exactly? If Bob wants to concede victory to the spooks, he needs a good scientific argument that Alice's alternative won't work. So far, Alice has the beginnings of an argument for the opposite conclusion. If we allow retrocausality, we can avoid spooks -- in the same way that superdeterminism does, but without abandoning free will. 

There are several lines of argument that Bob might try at this point. But before we turn to those, let's set aside another confusion. Some commentators have trouble seeing that retrocausality isn't just another version of action-at-a-distance. In other words, they have trouble seeing that it really does avoid the spooks. Let's try to get this straight.

\subsection*{\textbf{\emph{The Parisian zigzag}}}

The retrocausal proposal was first suggested by a young Parisian physicist, Olivier Costa de Beauregard, in the 1950s.\footnote{`M\'echanique quantique', \emph{Comptes Rendus Acad\'emie des Sciences} 236, 1632--34 (1953).}  He saw it as an objection to the famous Einstein-Podolsky-Rosen (EPR) argument from 1935. In that argument, EPR assume that there can be no direct action-at-a-distance, in order to conclude that a measurement on one particle cannot influence a distant particle. Costa de Beauregard pointed out that we could have distant influence without direct action-at-a-distance, so long as the influence takes an indirect ``zigzag'' path through space and time, via the point at which the two particles will have interacted in the past.

In the modern version of  the retrocausal proposal, when entangled particles are produced together, the zigzag works like this. Alice's (freely chosen) measurement setting affects her particle, back to the moment when the two particles were in contact, and hence affects Bob's particle, too, without having to jump any gap. So it affects Bob's measurement results, via a continuous path through spacetime.  In other words, Alice's choice of measurement setting does have a subtle influence on the results of Bob's measurements, away on the other side of the experiment. (The same holds in reverse: Bob's choice of settings has a subtle influence on Alice's results.) But it doesn't happen in the magical ``through thin air'' way that seemed so implausible to Einstein, Schr\"odinger and many others. It happens on a continuous path through spacetime, along the trajectories of the particles themselves.
 
In the new Hanson experiment, there are actually two pairs of particles, not just one.  Still, retrocausality can resolve the mystery in essentially the same way, operating on each separate pair.  Alice's influence is linked to a photon that goes to an intermediate location C, and Bob's influence is linked to another photon that also goes to C.\footnote{Arguably it is also natural to consider retrocausal influence from the intermediate measurement at C, but this additional retrocausality is not strictly required in this case.}  Given this extra information at C (available via the two zigzag paths) a selection device at that location can easily group some of the experiments into cases where the Alice-Bob correlations look ``spooky''.\footnote{This claim follows from the example of delayed-choice entanglement swapping (Ma, X.~\emph{et al,} `Experimental delayed-choice entanglement swapping', \emph{Nature Physics} 8, 480--485 (2012). \href{http://arxiv.org/abs/1203.4834}{arXiv:1203.4834}), which can be treated as an artifact of post-selection rather than any direct Alice-Bob connection. (We are grateful for comments from A.~Laidlaw at this point.)}

As Costa de Beauregard also pointed out, an attractive feature of his proposal is that because both arms of the zigzag lie in or on the lightcones, it is immediately congenial to special relativity, in a way in which direct action-at-a-distance is not. This also distinguishes retrocausality from other explanations that hope to fill the gap between entangled particles, perhaps with continuous processes travelling faster than light.

Many commentators think of the tension between action-at-a-distance and special relativity as the source of some of deepest puzzles about quantum theory. As David Albert and Rivka Galchen put it, in a piece in \emph{Scientific American:} ``Quantum mechanics has upended many an intuition, but none deeper than [locality]. And this particular upending carries with it a threat, as yet unresolved, to special relativity---a foundation stone of our 21st-century physics.''\footnote{Albert, D. Z.~and Galchen, R., `A quantum threat to special relativity', \emph{Scientific American} 300, 32--39 (2009)} But if we keep our eyes on the fact the zigzag retrocausal proposal avoids this tension completely, it is easy to see that it isn't just another form of spooky action-at-a-distance.

\subsection*{\textbf{\emph{Arguments against retrocausality?}}}

We've seen that retrocausality is different both from the direct action-at-a-distance usually held to be implied by the quantum correlations, and from superdeterminism. And it has advantages over both, in pretty obvious ways -- a better fit with relativity in one case, saving free will in the other. It sounds like a loophole worth noticing. Is there some good reason to rule it out of court?

Perhaps it leads to unacceptable paradoxes. Wouldn't backward causation lead to time travel-like contradictions? Modern Alice and Bob could signal to their medieval ancestors, sending them advice on contraception. But it is easy to show that the kind of subtle retrocausality needed to explain quantum correlations wouldn't have such consequences. It couldn't be used to signal for much the same reason that entanglement itself can't be used to signal.\footnote{Price, H.~and Wharton, K., `Disentangling the quantum world', \href{http://arxiv.org/abs/1508.01140}{arXiv:1508.01140}.}

Perhaps we should dismiss it because no one has been able to suggest a way in which it could be experimentally tested. But this, too, is a mistake. Retrocausality is proposed as an explanation of the correlations observed in all the standard tests of the relevant quantum  correlations, including those of the Hanson experiment. It is confirmed by those experiments to same extent that other proposed explanations are also confirmed -- including spooky action-at-a-distance itself. Experiments alone cannot distinguish between the various proposals. But this is no more a justification for ignoring retrocausality than it is for ignoring action-at-a-distance -- they are in exactly the same boat, as far as the experiments themselves are concerned, and the choice between them has to be made on other grounds.

Perhaps we should dismiss it because there is some deeper, philosophical objection to retrocausality. Ordinarily we think that the past is fixed, whereas the future is open, or partly so. Doesn't our freedom to affect the future depend on this openness? How could we affect what was already fixed? 

These are deep philosophical waters, but we don't have to paddle out very far to see that Alice has some options. She can say that according to the retrocausal proposal, quantum theory turns out to show that the division between what is fixed and open doesn't line up neatly with the distinction between past and future. Some of the past turns out to be open, too, in whatever sense the future is open. 

To decide what sense that is, precisely, we'd need to swim out a lot further. Is the openness Òthere in the worldÓ, or is it a projection of our own viewpoint as agents, making up our minds how to act? This really is a deep philosophical issue, but Alice doesn't need an answer to it. Whatever works for the future will work for the past, too, in the subtle way it needs to, to avoid the spooks. Either way, the result will be that our naive view of time needs to be revised a little bit, in the light of physics -- a surprising conclusion, perhaps, but hardly a revolutionary one, more than a century after special relativity wrought its own changes on the surrounding landscape. (Bob certainly needs a better argument for dismissing retrocausality than ``This isn't what we've always thought!'')

So far, Alice seems to have the better hand, and there are other high cards we haven't mentioned. For example, there are new arguments suggesting that time-symmetry might require retrocausality, for reasons novel to the quantum case.\footnote{Price, H., `Does time-symmetry imply retrocausality? How the quantum world says ``maybe''{'}, \emph{Studies in History and Philosophy of Modern Physics} 2012, 43, 75--83. Pusey, M., `Time-symmetric ontologies for quantum theory', presented at \emph{Free Will and Retrocausality in the Quantum World,} Trinity College, Cambridge, 3 July 2014 [\href{http://youtu.be/il19Unc5qVw}{http://youtu.be/il19Unc5qVw}]. For further discussion see Price, H.~\& Wharton, K., `Dispelling the Quantum Spooks -- a Clue that Einstein Missed?', \href{http://arxiv.org/abs/1307.7744}{arXiv:1307.7744}.} But our goal here is not to make a decisive case for retrocausality. We simply want to point out that in the absence of a good case against it, the argument for spooky action-at-a-distance cannot be considered settled.

Back in 1935, it seemed obvious to Einstein and others that there could be no action-at-a-distance. As Schr\"odinger put it at the time, commenting on the famous EPR paper, ``measurements on separated systems cannot directly influence each other -- that would be magic.''\footnote{\emph{Proceedings of the American Philosophical Society,} 124, 323--38. [Reprinted as Section I.11 of Part I of \emph{Quantum Theory and Measurement} (J.A. Wheeler and W.H. Zurek, eds., Princeton University Press, New Jersey 1983).]}  These days, the ``magic'' that Schr\"odinger derided in this way has become accepted orthodoxy. Worse still (from the Einstein-Schr\"odinger perspective), it is becoming orthodoxy that there is simply no alternative, at least unless we are prepared to embrace superdeterminism and abandon free will. It is too soon to tell whether the first orthodoxy will survive, but as things stand at the moment, the second is plainly false. The herd of mainstream opinion has resigned itself to the spooks, but there is an open gate leading in a different direction.

\end{document}